# Atom assisted cavity cooling of a micromechanical oscillator in the unresolved sideband regime

**Bijita Sarma and Amarendra K Sarma**

Department of Physics, Indian Institute of Technology Guwahati, Guwahati-781039, Assam, India
s.bijita@iitg.ernet.in, aksarma@iitg.ernet.in

**Abstract**. The ground state cooling of a mechanical oscillator in an optomechanical cavity containing an ensemble of identical two-level ground-state atoms is studied in the highly unresolved-sideband regime. The system exhibits electromagnetically-induced transparency-like quantum interference effect. The mutual interaction with the cavity optical field gives rise to an indirect coupling between the atomic and mechanical modes. In presence of this interaction, the noise spectrum gets modified and leads to asymmetric cooling and heating rates. Using the quantum master equation, time evolution of the average phonon number is studied and it is observed that the average phonon occupancy in the mechanical resonator exhibits ground-state cooling.

## 1. Introduction

Recently there has been a surge of interest in the study of micromechanical systems in the context of cavity optomechanics. Research in this field has been primarily carried out with a view towards the application of mechanical oscillators for ultrasensitive measurements and as tools for quantum information processing. Since many highly sensitive measurements are limited by thermal noise, it is a prerequisite to go down to the ground-state of these oscillators for these possible applications [1]. Thus ground-state cooling of micromechanical resonators has become a necessary prerequisite for all such studies. In analogy to the laser cooling of ions in strong binding regime [2], conventional cavity cooling of mechanical oscillators requires the condition of resolved-sideband regime, where the cavity mode decay rate should be lower than the mechanical oscillator resonance frequency. However, this condition is hard to be fulfilled in experiments. Here, we consider the ground state cooling of a micromechanical oscillator in an optomechanical cavity that traps an ensemble of identical two-level atoms. Atomic systems have drawn much curiosity in the context of application in the field of optomechanics due to the viability of coupling atoms to cavity and their comparatively lower decay rate [3]. We carry out the analysis in unresolved-sideband regime and we show that even in the highly unresolved-sideband regime, cavity cooling is possible by coupling the mechanical oscillator to the atomic ensemble via the cavity field. It is interesting to observe that this coupling can be harnessed to obtain unequal heating and cooling rates of the mechanical oscillator, thereby reaching the ground-state cooling.

## 2. Theory

We consider an optomechanical cavity containing an ensemble of identical two-level ground-state atoms as shown schematically in figure 1(a). The cavity is driven by an intense pump laser of frequency $\omega_l$. The Hamiltonian of the system reads

$$H = H_0 + H_I + H_{drive} + H_{bath} \qquad (1)$$

where, $H_0 = \omega_c c^\dagger c + \omega_m b^\dagger b + \omega_a S_z$ is the free Hamiltonian of the system, where $c$ and $b$ denote the bosonic annihilation operators of the cavity and mechanical modes, and $S_z = \sum_i \sigma_z^i$ is the collective z- spin operator of the atoms. The second term $H_I = g_{OM} c^\dagger c(b^\dagger + b) + J_0(S_- c^\dagger + S_+ c)$ describes the optomechanical and the atom-field interactions, where $g_{OM}$ is the single-photon optomechanical coupling strength and $J_0$ is the averaged atom-photon coupling strength. The coherent pumping of the optical mode is described by $H_{drive} = \varepsilon_p(c^\dagger e^{-i\omega_l t} + c e^{i\omega_l t})$ with amplitude $\varepsilon_p = \sqrt{\frac{k_c P_{in}}{\hbar \omega_l}}$. The last term, $H_{bath}$ describes the dissipations of the system. We transform the spin algebra of the atoms to a collective bosonic operator, $a = S_-/\sqrt{N}$. For large atom number and weak atom-photon coupling, $S_z \approx -N/2 + a^\dagger a$. In the frame rotating with the input laser frequency $\omega_l$, the Hamiltonian reads

$$H = -\Delta_c c^\dagger c - \Delta_a a^\dagger a + \omega_m b^\dagger b + g_{OM} c^\dagger c(b^\dagger + b) + J(c^\dagger a + a^\dagger c) + \varepsilon_p(c^\dagger + c) \qquad (2)$$

Here, $\Delta_c = \omega_l - \omega_c$ and $\Delta_a = \omega_l - \omega_a$ are the detunings of the cavity mode and the atomic mode with respect to the input laser. $J = J_0 \sqrt{N}$ is the collective atom-photon coupling strength. The time evolution of the system operators are described by nonlinear Heisenberg-Langevin equations

$$\dot{c} = \left(i\Delta_c - \frac{k_c}{2}\right) c - i g_{OM} c(b^\dagger + b) - iJa - i\varepsilon_p - \sqrt{k_c} c_{in}(t) \qquad (3.1)$$

$$\dot{a} = \left(i\Delta_a - \frac{k_a}{2}\right) a - iJc - \sqrt{k_a} a_{in}(t) \qquad (3.2)$$

$$\dot{b} = \left(-i\omega_m - \frac{\gamma_m}{2}\right) b - i g_{OM} c^\dagger c - \sqrt{\gamma_m} b_{in}(t) \qquad (3.3)$$

where, $k_c$, $k_a$ and $\gamma_m$ are the decay rates of the cavity mode, atomic mode and the mechanical mode respectively. $c_{in}$, $a_{in}$ and $b_{in}$ are the corresponding input vacuum noise operators with zero mean value and nonzero correlation functions given by $\langle c_{in}(t) c_{in}^\dagger(t') \rangle = \delta(t-t')$, $\langle a_{in}(t) a_{in}^\dagger(t') \rangle = \delta(t-t')$, $\langle b_{in}(t) b_{in}^\dagger(t') \rangle = (n_{th}+1)\delta(t-t')$ and $\langle b_{in}^\dagger(t) b_{in}(t') \rangle = n_{th} \delta(t-t')$ [4]. $n_{th}$ is the bath phonon number given by, $n_{th} = [exp\left(\frac{\hbar \omega_m}{k_B T}\right) - 1]^{-1}$, where $k_B$ is Boltzmann constant and $T$ is the bath temperature. For a strong input laser drive, equations (3.1)-(3.3) can be linearized around the steady state mean values as, $O = O_s + \delta O$. The Langevin equations for the fluctuation terms are given by

$$\delta \dot{c} = \left(i\Delta_c' - \frac{k_c}{2}\right) \delta c - iG(\delta b^\dagger + \delta b) - iJ\delta a - \sqrt{k_c} c_{in}(t) \qquad (4.1)$$

$$\delta \dot{a} = \left(i\Delta_a - \frac{k_a}{2}\right) \delta a - iJ\delta c - \sqrt{k_a} a_{in}(t) \qquad (4.2)$$

$$\delta \dot{b} = \left(-i\omega_m - \frac{\gamma_m}{2}\right) \delta b - iG(\delta c^\dagger + \delta c) - \sqrt{\gamma_m} b_{in}(t) \qquad (4.3)$$

where, $G = g_{OM} c_S$ is the enhanced optomechanical coupling strength due to the driving optical field and $\Delta_c' = \Delta_c - g_{OM}(b_s + b_s^\dagger)$ is the modified detuning. Then the linearized Hamiltonian of the system is given by

$$H_L = -\Delta_c' \delta c^\dagger \delta c - \Delta_a \delta a^\dagger \delta a + \omega_m \delta b^\dagger \delta b + G(\delta c^\dagger + \delta c)(\delta b^\dagger + \delta b) + J(\delta c^\dagger \delta a + \delta a^\dagger \delta c) \qquad (5)$$

The radiation pressure force on the mechanical oscillator for the hybrid system is estimated as $F = -\frac{\partial H_I}{\partial x} = -G[\delta c^\dagger + \delta c]/x_{ZPF}$, where, $x_{ZPF}$ is the zero-point fluctuation of the mechanical oscillator. The spectral density, for the system is calculated using $S_{FF}(\omega) = \int dt \, e^{i\omega t} \langle F(t) F(0) \rangle$, that equals to be

$$S_{FF}(\omega) = \frac{G^2 |\chi(\omega)|^2}{x_{ZPF}^2} [k_c + k_a J^2 |\chi_a(\omega)|^2] \qquad (6)$$

where, $\chi(\omega) = [\{\chi_c(\omega)\}^{-1} + J^2 \chi_a(\omega)]^{-1}$ is the total response function of the hybrid system. Response functions of the cavity optical mode and atomic modes are given by $\chi_c(\omega) = \left[-i(\omega + \Delta_c') + \frac{k_c}{2}\right]^{-1}$ and $\chi_a(\omega) = \left[-i(\omega + \Delta_a) + \frac{k_a}{2}\right]^{-1}$ respectively.

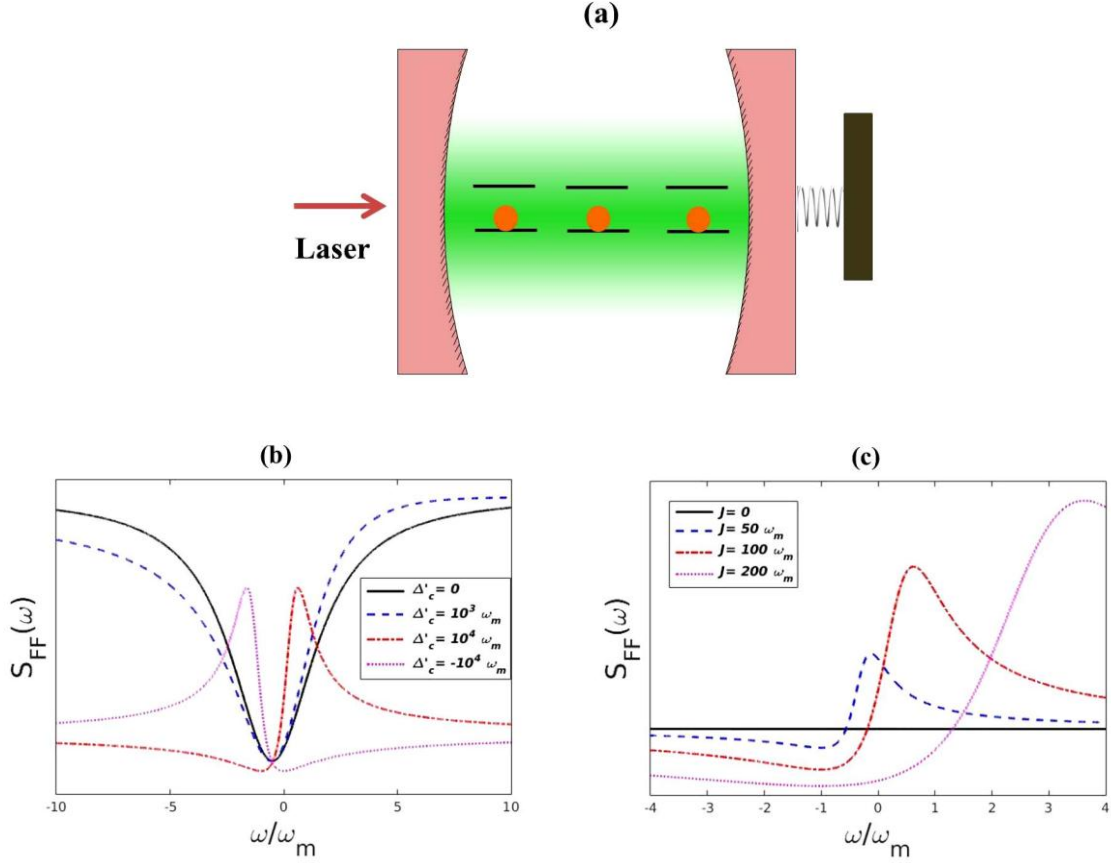

**Figure 1.** (a) Schematic diagram of an optomechanical cavity containing a two-level atomic ensemble; (b) plot of noise spectrum with varying $\Delta'_c$ for $J = 100\omega_m$; (c) plot of noise spectrum with varying $J$ for $\Delta'_c = 10^4\ \omega_m$. Other parameters are $\gamma_m = 10^{-5}\omega_m$, $k_c = 10^4\omega_m$, $k_a = 0.5\ \omega_m$, $\Delta_a = 0.5\ \omega_m$, $G = 50\omega_m$ and $n_{th} = 10^3$.

## 3. Results

In absence of the atomic ensemble, the setup reduces to a standard optomechanical system. In that case, optimum cooling occurs in the resolved sideband regime $k_c \ll \omega_m$ at cavity detuning $\Delta'_c = -\omega_m$. Here, we are considering the coherent interaction among the optical mode, the mechanical mode and the atomic mode. The optical mode is coupled to the mechanical mode via the optomechanical interaction and in turn the atomic mode is also connected to the optical mode. Thus these modes form a $\Lambda$ type three-level system. This type of systems exhibits quantum interference effect that is similar to the electromagnetically-induced transparency (EIT) in atomic systems. In figure 1(b), the noise spectrum for different values of modified cavity-field detuning in the unresolved-sideband regime is shown. For a generic optomechanical system, the noise spectrum is standard Lorentzian. However, in presence of the atomic ensemble, the spectrum is not Lorentzian. The cooling and heating rates of the mechanical resonator are given by $A_- = S_{FF}(\omega_m)x^2_{ZPF}$ and $A_+ = S_{FF}(-\omega_m)x^2_{ZPF}$ respectively. The plots clearly indicate that the cooling and heating rates are not equal. The large asymmetry between mechanical cooling and heating processes results in the possibility of enhancing the cooling rate while suppressing the heating rate. As can be seen from figure 1(b), the asymmetry in heating and cooling rates is prominent at $\Delta'_c = 10^4 \omega_m$ and $-10^4 \omega_m$. However, the detuning $\Delta'_c = 10^4 \omega_m$ would give rise to net cooling effect via the enhancement of cooling rate and inhibition of heating rate; whereas $\Delta'_c = -10^4 \omega_m$ would give rise to heating of the mechanical oscillator by the enhancement of heating rate and inhibition of cooling rate. Hence, by tuning the cavity-field detuning, the cooling rate of the mechanical resonator can potentially be enhanced while reducing the heating rate. For further calculations, detuning is taken to be $\Delta'_c = 10^4 \omega_m$. In figure 1(c), the variation of noise spectrum as a function of the atom-cavity coupling strength $J$ is shown. In the unresolved-sideband regime, in the absence of the atomic ensemble, the noise spectrum is flat as illustrated in the plot. For optimum value of $J$ near $100 \omega_m$, the asymmetry in cooling and heating rates is prominent.

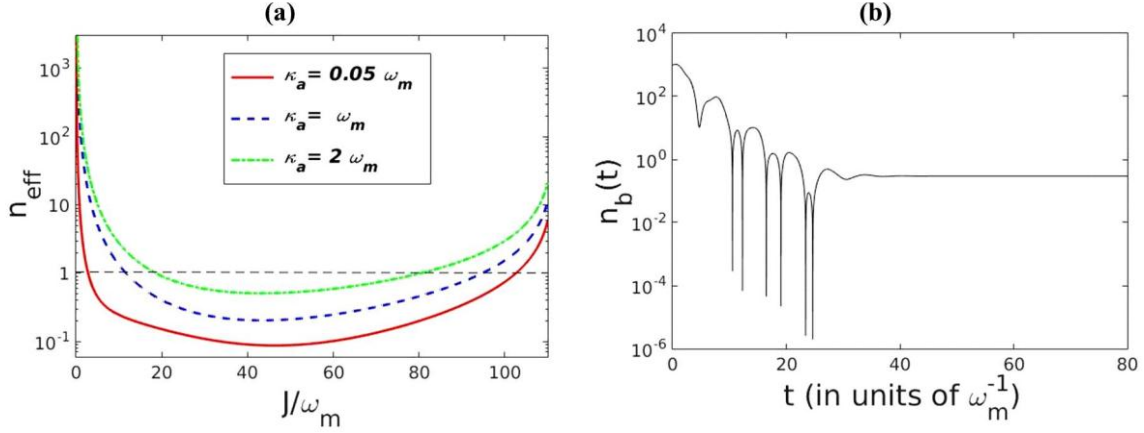

**Figure 2.** (a) Plot of steady-state cooling limit as function of $J$ for varying $k_a$; (b) Time evolution of the mean phonon number in the mechanical resonator in presence of the atomic ensemble, starting from $n_{th} = 10^3$. Parameters considered are $k_c = 10^4 \omega_m$, $k_a = 0.5 \omega_m$, $\Delta'_c = 10^4 \omega_m$, $\Delta_d = 0.5 \omega_m$, $\gamma_m = 10^{-5} \omega_m$, $J = 100 \omega_m$ and $G = 50 \omega_m$.

In the highly unresolved regime, $k_c \gg \omega_m$ and for $k_c \gg (k_a, \gamma_m)$, the cavity mode can be considered as a perturbation. Under this assumption, the three-mode system can be reduced to a two-mode system and in analogy to the standard cavity optomechanical system; the steady-state cooling limits are approximated as $n_{eff} = n_{classical} + n_{quantum}$. For effective resolved sideband ($k_{eff} \ll \omega_m$) and effective weak coupling regime ($G_{eff} < k_{eff}$) the classical and quantum cooling limits are given by $n_{classical} = \frac{4G_{eff}^2 + k_{eff}^2}{4G_{eff}^2 k_{eff}} \gamma_m n_{th} \approx \frac{k_{eff}}{4|G_{eff}|^2} \gamma_m n_{th}$ and $n_{quantum} = \frac{k_{eff}^2 + 8G_{eff}^2}{16(\omega_m^2 - 4G_{eff}^2)} \approx \frac{k_{eff}^2}{16\omega_m^2}$ [5]. The modified parameters for the effective atomic-mechanical mode interaction are given by: $\Delta_{eff} = \Delta_a - \eta^2 \Delta_c'$, $k_{eff} = k_a + \eta^2 k_c$, $G_{eff} = \eta G$, where $\eta = J/\sqrt{\Delta_c'^2 + \left(\frac{k_c}{2}\right)^2}$. Even when the cavity is in the unresolved-sideband regime, the effective interaction of the atomic and mechanical modes can bring the system back to effective resolved-sideband regime. In figure 2(a) we show the steady-state cooling limit $n_{eff}$ as a function of normalized atom-cavity coupling $J$, for different values of atomic line-width $k_a$. The plots show the broad range of coupling strength for optimum cooling. It is worth to be noted that for lower values of $k_a$, the cooling limit is lower and the range of $J$ is broader. Furthermore, we study the time evolution of the average phonon number $n_b(t) = \langle \delta b^\dagger \delta b (t) \rangle$ in the mechanical resonator, using the master equation approach. The quantum master equation for the system is given by

$$\dot{\rho} = i[\rho, H_L] + \frac{k_c}{2}(2\delta c\rho\delta c^\dagger - \delta c^\dagger \delta c\rho - \rho\delta c^\dagger \delta c) + \frac{k_a}{2}(2\delta a\rho\delta a^\dagger - \delta a^\dagger \delta a\rho - \rho\delta a^\dagger \delta a) +$$
$$\frac{\gamma_m}{2}(n_{th}+1)(2\delta b\rho\delta b^\dagger - \delta b^\dagger \delta b\rho - \rho\delta b^\dagger \delta b) + \frac{\gamma_m}{2}n_{th}(2\delta b^\dagger \rho\delta b - \delta b\delta b^\dagger \rho - \rho\delta b\delta b^\dagger) \quad (7)$$

To find out the mean phonon number $n_b(t)$, we solve a system of linear differential equations $\partial_t \langle \hat{o}_i \hat{o}_j \rangle = Tr(\dot{\rho}\hat{o}_i \hat{o}_j) = \sum_{m,n} \mu_{m,n} \langle \hat{o}_m \hat{o}_n \rangle$ where, $\hat{o}_i$, $\hat{o}_j$, $\hat{o}_m$, $\hat{o}_n$ are one of the operators: $\delta a^\dagger$, $\delta b^\dagger$, $\delta c^\dagger$, $\delta a$, $\delta b$ and $\delta c$. $\mu_{m,n}$ are the corresponding coefficients. In figure 2(b), the time evolution of the mean phonon number is shown. The bath phonon number is considered to be $10^3$. Initially the average phonon number in the mechanical oscillator is equal to the bath phonon number and all other second order moments are zero. The plot shows that the average phonon occupancy in the mechanical resonator tends to zero with time indicating ground state cooling of the resonator.

## 4. Conclusion

We have studied the cooling of a mechanical resonator in an optomechanical cavity containing an ensemble of two-level atoms in the highly unresolved-sideband regime. The system exhibits quantum interference effect that is similar to the electromagnetically-induced transparency (EIT) in Λ-type atomic systems. Though the atomic and mechanical modes are not coupled directly, the interaction with the cavity optical field gives rise to an indirect coupling between them. The noise spectrum gets modified due to quantum interference effect and leads to asymmetric cooling and heating rates. Using the quantum master equation, time evolution of the mean phonon number is studied and it is observed that the average phonon occupancy in the mechanical resonator exhibits ground-state cooling.